# Generative Steganography by Sampling

Zhuo Zhang, Jia Liu, Yan ke , Yu Lei, Jun Li, Minqing Zhang and Xiaoyuan Yang

e-mail: liujia1022@gmail.com

**ABSTRACT** In this paper, a novel data-driven information hiding scheme called "generative steganography by sampling" (GSS) is proposed. Unlike in traditional modification-based steganography, in our method the stego image is directly sampled by a powerful generator: no explicit cover is used. Both parties share a secret key used for message embedding and extraction. The Jensen-Shannon divergence is introduced as a new criterion for evaluating the security of generative steganography. Based on these principles, we propose a simple practical generative steganography method that uses semantic image inpainting. The message is written in advance to an uncorrupted region that needs to be retained in the corrupted image. Then, the corrupted image with the secret message is fed into a Generator trained by a generative adversarial network (GAN) for semantic completion. Message loss and prior loss terms are proposed for penalizing message extraction error and unrealistic stego image. In our design, we first train a generator whose training target is the generation of new data samples from the same distribution as that of existing training data. Next, for the trained generator, backpropagation to the message and prior loss are introduced to optimize the coding of the input noise data for the generator. The presented experiments demonstrate the potential of the proposed framework based on both qualitative and quantitative evaluations of the generated stego images.

## I. INTRODUCTION

In Fridrich's groundbreaking work of modern steganography[1], the possible steganographic channel are divided into three categories: cover selection, cover modification and cover synthesis. A cover selection method , likes an image retrieval method, retrieves stego images from real image database based on their fitness to carry a given secret image. In such a method, the cover image is not modified, it is not susceptible to existing steganalysis technology. However, such methods are not suitable for practical applications because of its low payload. Cover modification is the most studied class of methods to data. In terms of the Kullback–Leibler (KL) divergence as a security metric, such methods can achieve $\varepsilon$-security or perfect security for only certain explicit models. These methods, such as those presented in [2,3], face challenges in resisting steganalysis [4]. The third set of methods, cover synthesis seems more consistent with the earliest Cardan grille method [1]. In theory, cover synthesis is the most flexible and powerful because synthetic cover data can be adapted to the message of interest without restriction. In practice, however, cover synthesis is so difficult to automate that almost all known practical systems use modification-based methods instead.

Fortunately, a new class of data-based sampling tools called generative adversarial network (GAN) [5] has emerged that, is able to generate better synthetic images than previous generative models and has consequently become a new focus of research in artificial intelligence. The goal of a GAN is to estimate the potential distribution of existing data and then generate new data samples from the same distribution. A GAN consists of two neural networks, a generator and a discriminator, the generator tries to produce realistic samples that will fool the discriminator, while the discriminator tries to distinguish real samples from generated ones. Currently, GANs are widely used for computer vision [5], image synthesis [22], [23], natural language processing [13] and other areas, but are seldomly used in image steganography .

Inspired by Cardan grilles and GANs, in this paper, we will show that cover synthesis steganography can be viewed as a special case of image synthesis and will propose a new framework called generative steganography by sampling (GSS), in which the stego image is directly sampling by a well-trained image generator. Given a trained generative model, we search for the closest encoding of the stego image in the latent image manifold using a message loss and a prior losses, which are introduced as constraints on the sampling conditions to penalize message extraction error and unrealistic images. This encoding is then passed through the generative model to generate a carrier image.  Based on this new framework, a practical cover synthesis method is proposed that is based on a mask called digital Cardan grille and semantic image inpainting. The digital Cardan grille is used to determine the location in which to hide the message. The message is written in advance into an uncorrupted region that needs to be retained in the corrupted image based on the digital Cardan grille. Then, the corrupted image with the secret message is fed into a GAN to provide constraints for stego image generation. Back-propagation to the input noise vector is employed in our approach to find the encoding that is closest to the provided corrupted image. The constrained sampling process not only reconstructs the corrupted image but also generates a rational stego image with desired image content.

This paper has the following contributions:

1. We propose a data-driven framework for steganography based on training a generator with a large amount of data, thereby simplifying the design of steganography schemes. We treat cover synthesis steganography as a constrained image synthesis problem and take advantage of the recent advances in generative modeling. The message loss and the prior loss terms are proposed for penalizing message extraction error and unrealistic stego image. The steganographic process is largely automatable, and we can obtain a large number of stego images via sampling from the generator. This approach can also be extended to other media, such as text, audio and video.

2. We define a new criterion for steganographic security using the Jensen-Shannon (JS) divergence between the real-image distribution $p_{real}$ and the stego-image distribution $p_{stego}$ which used to measure how close these two distributions are to each other. Instead of ignoring the distribution of the real image, as in traditional cover modification steganography, we can define a random variable $z$ with a fixed distribution $p(z)$ and pass it through a parametric function $G_{stego}$ (typically a neural network of some kind) that directly generates stego image following a certain distribution $p_{stego}$. By using GAN, in the ideal case, $p_{stego} = p_{real}$ can be achieved, corresponding to statistically perfect stegnographic security.

3. Unlike texture-based methods, semantic image synthesis technology is used to ensure the rationality of the cover contents. Compared with our previous work, the new digital Cardan grille method proposed in this paper not only converges faster but also reduces visual distortion.

The remainder of this paper is organized as follows: Section II presents the related work on cover synthesis steganography. Section III describes the proposed method. Section IV presents the experimental results and analysis. Conclusions are presented in Section V.

## II. RELATED WORK

Fridrich, et al. [1] were the first to investigate a cover synthesis method. With the help of texture synthesis, the authors of [6, 7] used a texture sample and a set of color points generated based on a secret messages to construct a dense texture images. Qian, et al. [8, 9] proposed a robust method of steganography based on texture synthesis. Xue, et al. [10] used marbling, a unique texture synthesis method that allows users to deliver personalized messages hidden within beautiful, decorative textures. This type of texture-based steganography is based on the premise that the cover does not have to represent real-world content (which is counterintuitive for steganography, whose objective is typically to maintain the nature of the cover). Similar to [1], Zhou, et al. [11] proposed a cover selection method based on a bag-of-words model (BOW) in which a set of subimages with visual words related to the textual information is found. Then, the portions of the image containing these subimages are used as stego images for secret communication. Li et al.[36] propose a construction-based data hiding technique which transforms a secret message into a fingerprint image directly. They generates the fingerprint image based on a piece of hologram phase constructed from the secret message.

Recently, adversarial training has been applied to steganography. Volkhonskiy, et al. [12] proposed a new model for generating image-like containers based on a deep convolutional generative adversarial network (DCGAN) [13]. This approach generates embedded messages that are more secure against steganalysis using standard cover modification algorithms. Similar to [12], Shi, et al. [14] introduced a new GAN design with improved the convergence speed, training stability and image quality. Abadi [15] used adversarial training to teach two neural networks to encrypt a short message that would fool a discriminator. Tang, et al. [16] proposed an automatic steganographic distortion learning framework using a GAN composed of a steganographic generative subnetwork and a steganalytic discriminative subnetwork. However, most of these GAN-based steganographic schemes still function based on the principle of cover modification. Similar to [15], the authors of [17] defined a game among three parties—Alice, Bob and Eve—to simultaneously train both a steganographic algorithm and a steganalyzer. However, this scheme still trains Alice to learn to produce a steganographic image using a modification scheme. These methods focus on the adversarial game while ignoring the core aim of the GAN—to build a powerful sampler.

Because a GAN's greatest advantage is sample generation, it seems intuitive to use a GAN to generate a semantic carrier image directly. However, the question of how to embed secret message in an image generated by GAN is a key issue. Some researchers have made preliminary attempts to apply this intuitive idea. Ke et al.[18] first proposed a generative steganography method called GSK, in which the secret messages are embedded into a generated cover image rather than an existing cover, thus resulting in no modifications in the cover. In [18], the term "generative steganography" was first introduced. Liu, et al. [19] proposed a method in which an auxiliary classifier GAN (ACGAN) [20] is applied to classify the generated samples and the class output information is used as the secret message. Hu, et al.[31] propose a method called steganography without embedding (SWE), in which the secret information is mapped to a noise vector and trained neural network model is used as a generator to generate a carrier image based on the noise vector. In the method proposed in our previous work [21], the secret message is written into the corrupted area of a corrupted image that needs to be filled in a grille mask; then, the corrupted stego image is fed into a GAN for stego generation.

## III. GENERATIVE STEGANOGRAPHY FRAMEWORK

The generative steganography framework proposed in this paper is illustrated in Fig. 1. In this scenario, the sender directly creates a stego carrier using a generator with a message and a key. Here, the embedding algorithm becomes a stego sampling (generation) process. The secret key shared by both parties ensures the security of message extraction, and the degree of naturalness of the stego image determines the security of the communication channel.

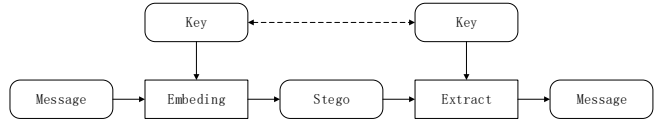

**FIGURE 1. Generative steganography framework.**

### A. REPRESENTATION FORMULA

Ideally, a generative steganography scheme should satisfy the following three conditions, which we call the generative steganography conditions (GSC):

$$s = \text{Gen}(m,k), \tag{1}$$

$$m = \text{Ext}(s,k), \tag{2}$$

$$p_{stego} = p_{real}, \tag{3}$$

where $s$ denotes the fake stego carrier, $m$ denotes the message, and $k$ is the secret information shared by the two parties. Eq. (1) represents the embedding procedure, Gen(·) is a generator, indicating that the scheme involves generative steganography. Note that the stego image $s$ is sampled by the generator without requiring an explicit cover image. Ext(·) in Eq.(2) denotes the message extraction operation. In Eq.(3), $p_{stego}$ and $p_{real}$ denote the distributions of the sampled stego carrier and the real data, respectively. Ideally, $p_{stego}=p_{real}$, this condition guarantees the statistical security of the communication channel, which is a key issue for a generator to be used for generative steganography. There are two main tasks in generative steganography. One is to construct a generator that can generate stego images such that the generated stego samples are subject to the same distribution as the real image distribution. Second, a message extraction algorithm must be designed. In this paper, we use a GAN to achieve the first goal. We also constrain the generation process to simplify the design of the message extraction algorithm. Interestingly, if we assume that all these conditions are satisfied, then generative steganography can be treated as a cover selection method in which it is possible to construct a fake database of infinite size in which every sample contains a secret message.

### B. A MEASURE OF SECURITY

We are motivated by the notion of $\varepsilon$-secure steganography proposed by Cachin [30], in which the relative entropy (also called the KL divergence) between the cover and stego distributions is less than or equal to $\varepsilon$. The KL divergence measures how one probability distribution diverges from another expected probability distribution. Fridrich [1] also introduced a formal information theoretic definition of security in steganography based on the KL divergence between the distributions of the cover and stego objects:

$$D_{KL}(p_{stego} \| p_{cover}) = E_{p_{cover}} \left[ \log \frac{p_{stego}}{p_{cover}} \right], \quad (4)$$

where $p_{cover}$ and $p_{stego}$ are the cover and stego distributions, respectively. However, the KL divergence does not satisfy the symmetric and triangle inequality conditions, and it cannot be strictly considered a metric. Thus, the security of different steganographic images cannot be evaluated in terms of this divergence. In this paper, a new measure of security is defined using the Jensen-Shannon (JS) divergence, which is based on the KL divergence but exhibits several notable (and useful) differences, including that it is symmetric and always has a finite value. The JS divergence is defined as follows:

$$D_{JS}(p_{stego} \| p_{cover}) = \frac{1}{2} D_{KL}(p_{stego} \| M) + \frac{1}{2} D_{KL}(p_{cover} \| M) \quad (5)$$

$$M = \frac{1}{2}(p_{stego} + p_{cover}). \quad (6)$$

The JS divergence between two probability distributions is bounded by 1 if the base 2 logarithm:

$$0 \le D_{JS}(p_{stego} \| p_{cover}) \le 1. \quad (7)$$

When $p_{stego} = p_{cover}$, the JS divergence is zero.

In generative steganography, we can use this metric to evaluate which generator is closest to a real data distribution. This means that we can sample a secure stego image using the best generator. In fact, the generator used in the study that introduced the concept of GANs [5] was trained based on the JS divergence; in this case, the adversarial game gradually reduces the divergence between the generator distribution $p_g$ and the data distribution $p_{data}$ with an increasing number of adversarial iterations. In generative steganography, the stego image is generated by the generator $G$. Therefore, $p_g$ and $p_{stego}$ have same meanings, and $p_{cover}$ is the same as $p_{data}$. Because no explicit cover exists in our scheme, we use $p_{data}$ instead of $p_{cover}$.

Similar to Cachin's $\varepsilon$-security for steganography [30], we define the concept of $\varepsilon$- security for generative steganography system based on the JS divergence :

$$D_{JS}(p_{stego} \| p_{data}) \le \varepsilon \quad (8)$$

When introducing the GAN concept[5], Goodfellow et al. proved that if generator $G$ and discriminator $D$ have sufficient capacity, the discriminator can reach its optimum given $G$, and $p_g$ is updated to improve its generation ability, then $p_g$ converges to $p_{data}$, which also means that $p_{stego}$ converges to $p_{data}$. Ideally, when the generator is optimal (i.e., $\varepsilon = 0$) the system can be regarded as perfectly statistically secure against steganalysis.

Under the generative steganography framework, we want find approximations to the real image distribution $p_{data}$ (which is complex and intractable) in the form of a (tractable) approximate distribution $p_{stego}$. In KL divergence, the term $p_{stego}\log[ p_{stego}/p_{data}]$ appears in the integrand. When $p_{data}$ is small, this divergence grows very rapidly if $p_{stego}$ is not also small. Thus, if one is choosing $p_{stego}$ to minimize $D_{KL}(p_{stego}//p_{data})$, it's very improbable that $p_{stego}$ will assign much mass on regions where $p_{data}$ is near zero. By contrast, the JS divergence does not have this property. It is well behaved when both $p_{data}$ and $p_{stego}$ are small. This means that it will not apply as strong penalty to a distribution $p_{stego}$ from which it is possible to sample values that are impossible in $p_{data}$. Admittedly, the JS divergence is also not an ideal choice. If two distributions $p_{data}$ and $p_{stego}$ are so far apart that there is no overlap at all, then the KL divergence is meaningless, whereas the JS divergence has a constant value. This behavior is essentially fatal in gradient-based learning algorithms because it means that the gradient at this point is zero. Arjovsky proposed an alternative, called Wasserstein GAN (WGAN) [32], which leverages the Wasserstein distance to produce a value function with better theoretical properties than the JS divergence. However, since a DCGAN is used to generate stego images later in this paper and the JS divergence is used in DCGAN, we use the JS divergence as our standard for security evaluation in this paper. Nevertheless, it should be noted that this is not the only option in the context of generative steganography frameworks.

*C. CONSTRAINED SAMPLING*

Currently, the state-of-the-art methods of cover modification steganography can be viewed as a constrained coding problem of minimizing the distortion between the cover and stego images through syndrome trellis coding (STC) [4]. The embedding and extraction mappings are realized using a binary linear code *C* of length *n* and dimension *n* - *m*:

$$\text{Ext}(\text{Emb}(\boldsymbol{x},\boldsymbol{m})) = \boldsymbol{m}, \ \forall \boldsymbol{x} \in \{0,1\}^n, \boldsymbol{m} \in \{0,1\}^m, \quad (9)$$

$$\text{Emb}(\boldsymbol{x},\boldsymbol{m}) = \underset{\boldsymbol{y} \in \zeta(\boldsymbol{m})}{\arg\min} D(\boldsymbol{x},\boldsymbol{y}), \quad (10)$$

$$\text{Ext}(\boldsymbol{y}) = \text{H}\boldsymbol{y}, \quad (11)$$

where *x* is the cover, *m* denotes the message, and *y* is the stego image. $D(\boldsymbol{x}, \boldsymbol{y})$ is the distortion function. H is a parity-check matrix of the code $\zeta$, and $\zeta(\boldsymbol{m})$ is the coset corresponding to syndrome *m*. Emb(.) and Ext(.) denote the embedding and extraction operations, respectively. The embedding operation involves finding an optimal solution to the problem of finding a stego image *y* that satisfies the message extraction conditions while simultaneously minimizing the distortion. The Viterbi algorithm can provide an optimal solution to the embedding problem. Such steganography implementations that lack a shared secret are forms of security through obscurity, which relies on the secrecy of either the design or the implementation as the main approach for providing security for a system or a component of a system.

In our generative steganography method, cover synthesis steganography is treated as a constrained image generation problem, and we take advantage of recent advances in generative modeling. Here, we provide a representation of the optimization problem for generative steganography:

$$\text{Ext}(\text{Gen}(\boldsymbol{m},\boldsymbol{k}),\boldsymbol{k}) = \boldsymbol{m}, \ \forall \boldsymbol{m} \in \{0,1\}^m, \quad (12)$$

$$\text{Gen}(m,k) = \underset{y \sim p_{stego}}{\arg\min} \text{D}_{\text{JS}}(p_{stego}, p_{data}), \quad (13)$$

$$\text{Ext}(\boldsymbol{y},\boldsymbol{k}) = \text{C}_k \boldsymbol{y}, \quad (14)$$

where Gen(.) is a generator trained by a GAN, and $C_k$ is an extraction matrix based on the secret key *k*. Importantly, unlike in the STC method, no explicit cover *x* is necessary as shown in Eq.(13). The stego image *y* does not depend on any specific cover; instead, it can be regarded as a sample from a generator distribution $p_{stego}$. Rather than estimating the density of $p_{data}$, which may not exist, the GAN defines a random variable *z* with a fixed distribution $p_z$ and pass it through a parametric generator $G_\theta$ that directly generates samples $y = G_\theta(z)$ following a certain distribution $p_{stego}$. By varying $\theta$, we can modify this distribution to bring it closer to the real data distribution $p_{data}$. Note that our scheme is also key-dependent and adheres to Kerckhoff's principle. In the next section, we will present a practical algorithm that reveals the details of $C_k$. In generative steganography, we not only require the JS divergence between $p_{stego}$ and $p_{data}$ to be as small as possible, as shown in Eq.(13), but also require the generator to satisfy the request for message extraction, as shown in Eq. (14).

In our constrained image generation scheme, the generative steganography procedure is implemented in two steps. First, we train a generator with GAN, as shown in Fig. 2(a). As discussed in Section III.B, the training target is to minimize $D_{JS}(p_{stego}, p_{data})$. Ideally, the training process will reach the state in which $p_{stego} = p_{data}$.

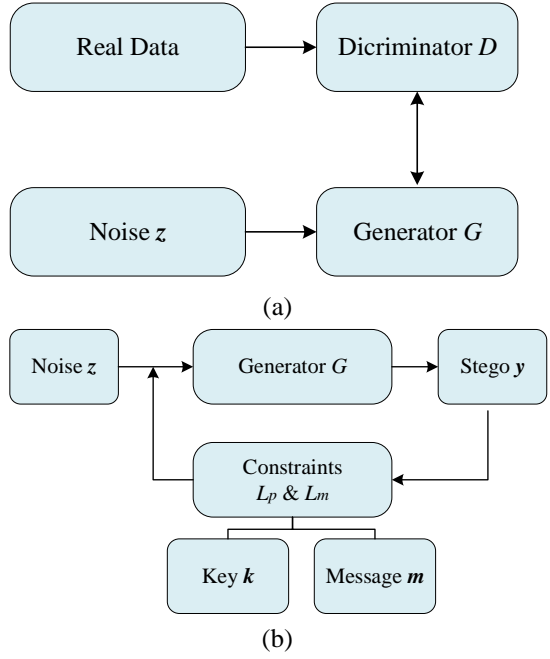

FIGURE 2. Constrained stego sampling (a) training a generator with a GAN; (b) sampling a stego image with constraints.

Next, we use the trained generator to search for stego images. More specifically, we formulate the process of finding the stego image *y* as a constrained sampling problem. The objective of this problem is to generate the message carrier directly from the noise *z* via generator. This secret carrier needs to satisfy two constraints: the stego image must be sufficiently realistic, and the stego image needs to contain the desired messages. In this paper, we use these two constraints to update and sample *z*, to achieve the purpose of generating the secret carrier. Let *m* be the message, and let *k* be a secret key shared by the two communicating parties. Using this notation, we define the "closest" encoding $\hat{z}$ as follows:

$$\hat{z} = \arg\min_{z}(L_m(z\mid m,k) + \lambda L_p(z)), \qquad (15)$$

where $L_m$ denotes the message loss, which constrains the generated image given the message *m* and the extraction key *k*; $L_p$ denotes the prior loss, which penalizes unrealistic images; and $\lambda$ is a hyperparameter that controls the importance of the prior loss relative to the message loss, as shown in Fig. 2(b). Backpropagation to the input data is introduced to optimize the coding of the input data *z* for the DCGAN:

$$z \leftarrow z - \gamma_z \nabla_z L, \qquad (16)$$

$$\nabla_z L = -\frac{\partial(L_m + \lambda L_p)}{\partial z}. \qquad (17)$$

As Eq. (16) and (17) show, we iteratively update *z* using backpropagation. After sufficient training iterations, the input data *z* for the GANs are optimized to minimize the loss. We obtain the generated stego image $y = G(\hat{z})$. The details of the proposed loss function are discussed in the following sections.

## IV. GENERATIVE STEGANOGRAPHY BY INPAINTING

Semantic inpainting [22] refers to the task of inferring arbitrarily large missing regions in images based on image semantics. Since high-level context prediction is required, this task is significantly more difficult than classical inpainting or image completion which is often more concerned with correcting spurious data corruption or removing entire objects. As far as we know, we are the first to note the similarity between image completion and Cardan-grille-based secret writing. Under the guidance of this constrained sampling framework, inspired by image inpainting and the Cardan grille approach, we propose a practical generative steganography method based on semantic inpainting.

The proposed steganography method consists of four phases, as illustrated in Fig. 3. In the first phase, we train a DCGAN on an image set and obtain a generator $G$ after DCGAN convergence. After training, the generator $G$ is able to take a point $z$ drawn from $p_z$ and generate an image that mimics samples from $p_{data}$. In the second phase, a constraint sampling condition is established by embedding a secret message into a corrupted image. In this paper, the message is written into the uncorrupted image region that must be preserved in the corrupted image through bit-plane embedding; the corrupted stego image will play an important role in the next phase, in which the purpose is to find natural stego images though iterative sampling. In the third phase, using the defined prior and context losses for completion and the message loss for steganography, the corrupted stego image can be mapped to the closest $z$ in the latent representation space, which we denote by $\hat{z}$. In this study, **z** is randomly initialized and updated using backpropagation on the total loss. The process of information hiding is consistent with the basic idea of the traditional Cardan grille: the sender defines a mask, called a digital Cardan grille, to determine where the message is hidden, and $\odot$ denotes the elementwise product operation between two data matrices. In the last phase, the stego image **y** can be easily obtained as $y = G(\hat{z})$, and a well-filled stego image is transmitted to the recipient through a public channel. The receiver extracts the secret message in the reconstructed image using the Cardan grille shared by the two parties. The core requirement for this method is to define a generator that can ensure not only the consistency of the secret messages but also the natural appearance of the stego images. In fact, it is feasible to use noise to generate images directly without using image completion technology, but the advantage of image completion is that we can control more of the details of image generation, giving the sender more selectivity, instead of giving the sampler all the selectivity.

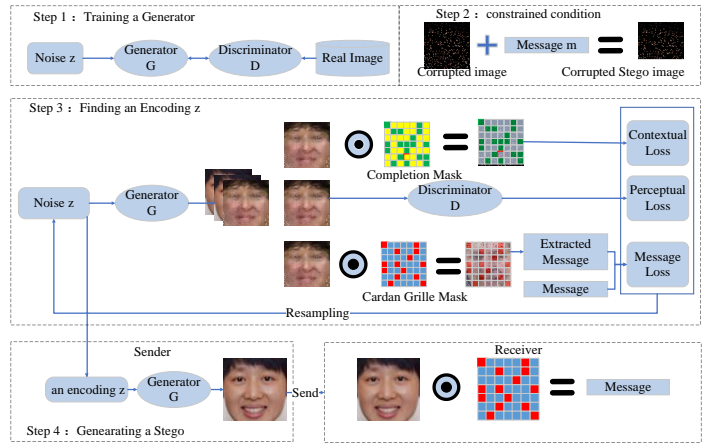

**FIGURE 3**. The proposed generative steganography by sampling method using the Cardan grille.

Before constructing the practical generative steganography algorithm, we first assume that the generator meet the condition $D_{JS}(p_{stego}, p_{data}) = 0$ or $D_{JS}(p_{stego}, p_{data}) < \varepsilon$; This condition can be reached by training a generator with a GAN as shown in step1 in Fig.3 Then, we can focus on how to design a scheme to ensure that the messages can be extracted correctly. In this paper, the process of sampling the stego image is decomposed into three steps to simplify the design.

*A. MESSAGE PREPROCESSING*

The objective of message preprocessing is to establish constrained sampling conditions. We define an expansion operation called *Expand*(.):

$$m' = Expand(m). \qquad (18)$$

$m$ is the original secret message, and $m'$ is a semifinished stego carrier containing the secret message $m$. Secret message $m$ can be extracted via the message extraction operation with the extraction key $k$:

$$m = \text{Ext}(m', k). \qquad (19)$$

In this paper, we construct a semifinished stego carrier based on corrupted image in image completion.

First, we select the corrupted image $I_{corrupted}$, which has dimensions of $m \times n$, along with the secret message $m$ and the Cardan grille $C_k$. $C_k$ is a key shared by both parties. A Cardan grille with the same size is defined as follows:

$$C_k = \begin{bmatrix} c_{11} & \cdots & c_{1n} \\ \vdots & \ddots & \vdots \\ c_{m1} & \cdots & c_{mn} \end{bmatrix}_{m \times n}, \qquad (20)$$

where $c_{ij} \in \{0,1\}$, A value of 1 represents a parts of the image where we want to hide the message, and a value of 0 represents a part of the image that we cannot embed to. Ideally, the Cardan grille should be designed to have a $m \times n$ bit key, where the key length coincides with the lower bound on the algorithm's security. It is important to note that the structure and location of this Cardan grille in the corrupted image are shared by both parties.

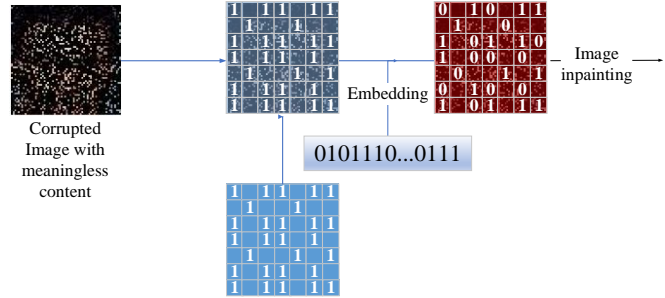

**FIGURE 4**. Flowchart of building the constraint conditions.

Then, the message can be written into the uncorrupted regions of the input image in accordance with the Cardan grille, as shown in Fig. 4. We obtain a corrupted image $I_{corrupted\_stego}$ which containing the secret message (denoted by $m'$). Note that $m = m' \odot C_k$, where $\odot$ denotes the elementwise product operation. The preprocessing is important because it transforms the image completion process into steganography. Then, we can obtain the final stego carrier using

$$s = \text{Gen}(m', k). \qquad (21)$$

In the next subsection, we will present details on how to realize the Gen() operation based on the GAN approach and image completion, thus completing the generative steganography procedure.

*B. FINDING AN ENCODING.*

As mentioned above, an image completion method used for steganography should satisfy two objectives: one is to preserve the rationality of the complete image content and the other is to provide a stable message. In this paper, we use the image inpainting method proposed by Yeh [22] based on DCGAN [13] for image completion. A binary mask *M*, which contains values of 0 or 1, is used for image completion. A value of 1 represents a part of the image that we want to keep, and a value of 0 represents a part of the image that we want to complete. To complete the image while hiding the message, we designed three losses.

**Contextual Loss**: To fill in large missing regions, similar to Ye's method [22], our method takes advantage of the remaining available data. To maintain the same context as in the input image, we want to ensure that the known pixel locations in the input image $I_{corrupted\_stego}$ will be similar to the pixels in $G(\mathbf{z})$. We need to penalize $G(\mathbf{z})$ when it does not create a similar image with respect to the pixels we know about. Formally, we do this by subtracting the pixels in $I_{corrupted\_stego}$ from $G(\mathbf{z})$ in an elementwise fashion and assessing how much they differ:

$$L_{contextual}(z) = \left\| M \odot G(z) - M \odot I_{corrupted\_stego} \right\|_1, \quad (22)$$

where $\|.\|_1$ is the $L_1$-norm. In the ideal case, all the pixels at known locations will be identical between $I_{corrupted\_stego}$ and $G(\mathbf{z})$. In this case, $G(\mathbf{z})_i - (I_{corrupted\_stego})_i = 0$ for the known pixels $i$; thus, $L_{contextual}(\mathbf{z}) = 0$.

**Perceptual Loss**: For an image to be considered to have a realistic appearance, the discriminator must be convinced that the image is real. We test this condition using the same criterion used when training the DCGAN:

$$L_{perceptual}(z) = \log(1 - D(G(z))). \quad (23)$$

The contextual loss and the perceptual loss successfully predict the semantic information in the missing region and achieve pixel-level photorealism.

**Message Loss**: The key to using image completion for information hiding is that the messages extracted with the Cardan mask $C_k$ should be as stable as possible. The pixel values at the corresponding positions in the generated image should be equal to the values in the secret message:

$$L_{message}(z) = \left\| [C_k \odot G(z)]_{BPI} - [C_k \odot m']_{BPI} \right\|_1 \\ = \left\| [C_k \odot G(z)]_{BPI} - m \right\|_1 \quad (24)$$

In the ideal case, all the pixels at the hiding locations should be identical between **m**' and $G(\mathbf{z})$. In this case, $G(\mathbf{z})_i - \mathbf{m}'_i = 0$ for the known pixels $i$; thus, $L_{message}(\mathbf{z}) = 0$. The notation $[\cdot]_{BPI}$ indicates that the steganography in performed in a bit plane. We define a bit plane index *BPI* (*BPI* = 1,..., 8.) to indicate the layer in which the message is located, *BPI* = 1 represents the least significant bit (LSB), and *BPI* = 8 represents the most significant bit (MSB). In the case that $I_{corrupted\_stego} = \mathbf{m}'$, $C_k = M$. $L_{message}$ is the same as $L_{contextual}$. It is important to note that $M$ and $C_k$ play different roles in stego image generation: $M$ is used for image inpainting, and $C_k$ is used for hiding information. The size and value of $C_k$ can differ from those of $M$. In practice, for each 8-bit pixel point in each layer, we cannot guarantee that the generator will converge to a model that can successfully satisfy $L_{message}(\mathbf{z}) = 0$. Intuitively, we expect that the lower bits will be more strongly affected by pixel generation, while higher bits will have greater stability. The elementwise product $\odot$ operates at the bit-plane level.

Finally, we can find $\hat{z}$ using a combination of all these losses as follows:

$$L(z) = L_{perceptual}(z) + L_{contextual}(z) + \lambda L_{message}(z), \quad (25)$$

$$\hat{z} = \arg\min L(z), \quad (26)$$

where $\lambda$ is a hyperparameter that controls the importance of the message loss relative to the perceptual loss and contextual loss. Back-propagation to the input noise $z$ is employed in our approach to find the encoding that is closest to the provided corrupted stego image.

*C. MESSAGE EXTRACTION*

For the message receiver, message extraction is simple. The basic process is shown in Fig. 5.

The receiver places the grille directly on the generated image to extract the secret message in the corresponding position. The extraction operation is

$$\boldsymbol{m} = [G(z) \odot C_k]_{BPI}. \quad (27)$$

The notation $[\cdot]_{BPI}$ indicates in which bit plane the extraction operation is performed.

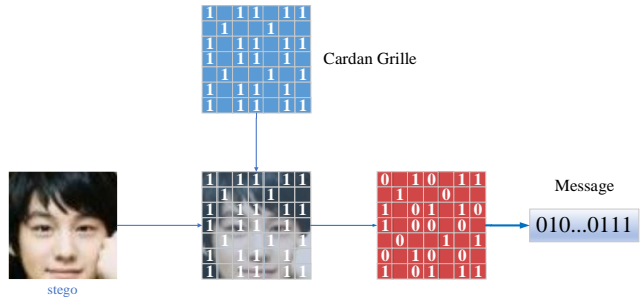

**FIGURE 5. Message extraction using a Cardan grille**

## V. EXPERIMENTS

In this section, we evaluate our experimental results both qualitatively and quantitatively.

*A. DATASETS AND SETTINGS*

We implemented our adversarial training scheme on the LFW dataset [23], a database of face photographs designed for studying the problem of unconstrained facial recognition. This dataset contains more than 13,000 images of faces collected from the web. We also use the CelebFaces Attributes (CelebA) dataset which is a large-scale face attributes dataset with more than 200k celebrity images, each with 40 attribute annotations to train the generator. For these two databases, we used an alignment tool to resize the images to 64×64 pixels, as shown in Fig. 6. The DCGAN model architecture used in this work was adopted from Yeh et al. [22]. We modified Brandon Amos's implementation [24] for hiding information. We used 12,000 samples from LFW and 150,000 samples from CelebA to train the DCGAN and set the training parameters for image completion to the same values used by Brandon Amos. The generator $G$, takes a random 100-dimensional vector drawn from a uniform distribution in the range [-1; 1] and generates a $64 \times 64 \times 3$ image. The discriminator model, $D$, is structured essentially in the reverse order. The input layer accepts a $64 \times 64 \times 3$ image, and it is followed by a series of convolution layers; in each of these layers, the image dimensions diminish by half, but the number of channels is twice that in the previous layer. The output layer is a two-class softmax layer. We set $\lambda = 0.1$ in our experiments. In the stego generation stage, we need to find $z$ using backpropagation. We used the Adam_algorithm [35] for optimization and restricted $z$ to [-1; 1] in each iteration because we found that this setting produced more stable results. We terminated the back-propagation process 1,000 iterations. We used identical settings for all tested datasets. The grille size was fixed to 64×64—the same size as the corrupted image. We intentionally randomized the secret message in all the uncorrupted regions in order to demonstrate the stability of the embedded messages in a quantitative manner. In the training phase, the model was run on a machine with an NVIDIA TiTan Xp GPU and 32GB of memory. We trained the DCGAN for 200 epochs on each ot the two datasets, CelebA and LWF, for 2 days and 1 day, respectively.

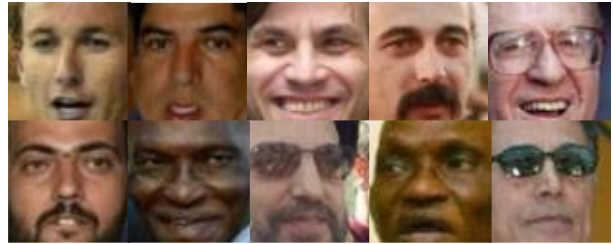

(a) LWF

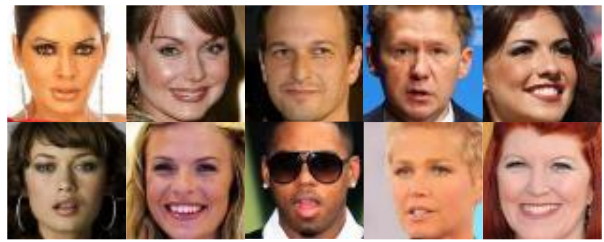

(b) CelebA

FIGURE 6. Aligned samples from the (a) LFW and (b) CelebA datasets

We tested three random pattern masks with different mask shapes: 1) random pattern masks with approximately 20% missing, 2) randomly horizontal or vertical masks with 50% missing, and 3) completely random masks with 90% missing.

## B. VISUAL COMPARISONS

Our results are shown in Fig. 7 and illustratie that our method can successfully predict the missing content given different random masks on the LWF and CelebA datasets. It is important to emphasize that in our experiment, the Cardan grille was randomly generated, and into all the locations available for hidden content, we wrote a message that was also randomly generated. Note that stego image generation (semantic inpainting in this case) is not an attempt to reconstruct the ground-truth image; instead, the goal is to fill in the holes with realistic content while hiding information. Even the ground-truth image is only one of many possibilities.

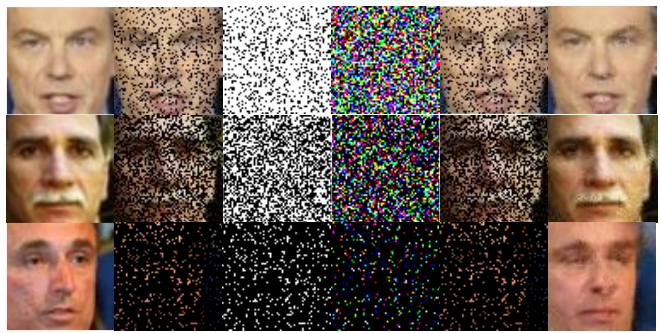

(a) LWF

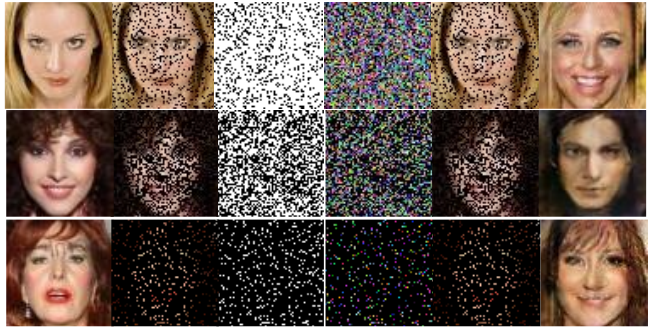

**(b) CelebA**

**FIGURE 7.** For each example, column 1 shows the ground-truth image from the dataset, column 2 shows an images with random regions missing (row 1: 20%, row2: 50%, row3: 90%), column 3 shows Cardan grille mask, column 4 shows secret message, column 5 shows corrupted stego image with BPI = 1 and column 6 shows the inpainting results obtained for the image in column 5 using our method.

We show the complete image generation process in Fig. 8, where the number of iterations ranges from 20 to 2,000. We samples 8 generated stego images from the generator. Note that we the ground-truth images were chosen from the LFW dataset in Fig. 8 (a) and from CelebA dataset in Fig. 8 (b). As shown in Fig. 8, meaningless severely corrupted images (with 90% missing) canbe transformed into samples consisten with $p_g$. In the first few rounds, the visual quality of the generator output is low. The image becomes increasingly realistic as the number of iterations increases. At the same time, we can also see that the generated image quality achieved based on the CelebA database is better than the result of training using LWF database due to the use of more samples for training. It can also be seen from Fig. 8(b) that for corrupted images with different contents, In the early stage of iterative sampling, the images may be very similar, as in the cased of row 1 and row3. However, due to the use of different constraints, the images tend to have increasingly different contents as the number of iteration increases.

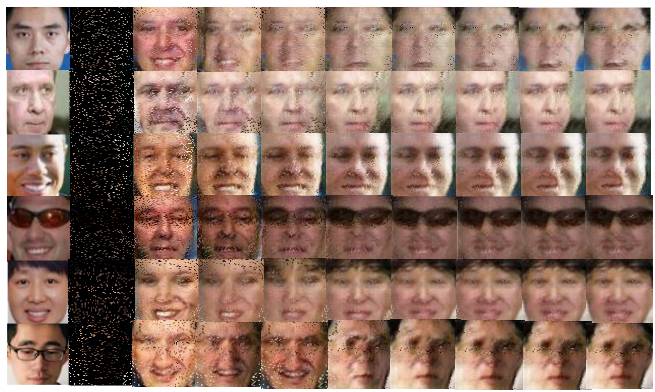

**(a) LWF**

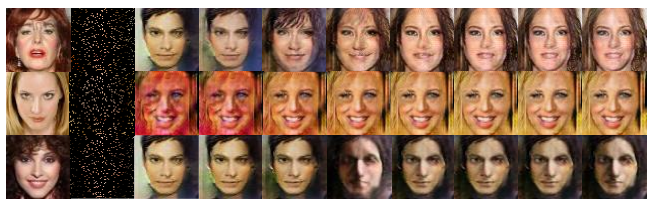

**(b) CelebA**

**FIGURE 8.** For each example, column 1 shows ground-truth image from the dataset; column 2 shows corrupted stego images which is missing a random regions comprising 90% of the image, and the Columns 3–10 show samples from the generator as the number of iterations increases.

In Fig. 9, panels (a) and (b) show the message loss and the perceptual loss, respectively, for single image. All the stego images were sampled at 500 iterations from corrupted images with missing regions comprising 90% of the image. As seen in Fig. 9(a), during the first few sampling rounds, the visual quality of the output is low and the perceptual loss is high; however, after approximately 250 iterations, the perceptual loss causes the generated sample to become increasingly realistic and natural. As seen in Fig. 9 (b), the message loss becomes relatively smooth and stable after 150 iterations. This is mainly because we set the message loss to have more influence on the total loss.

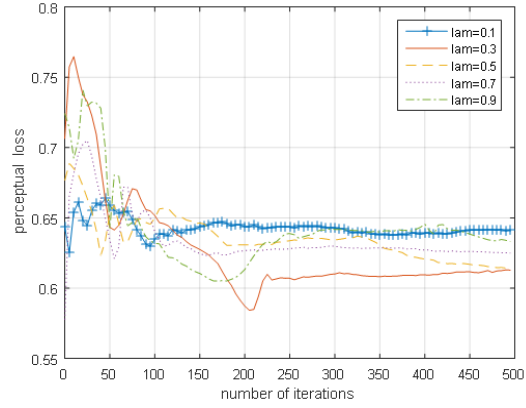

(a)

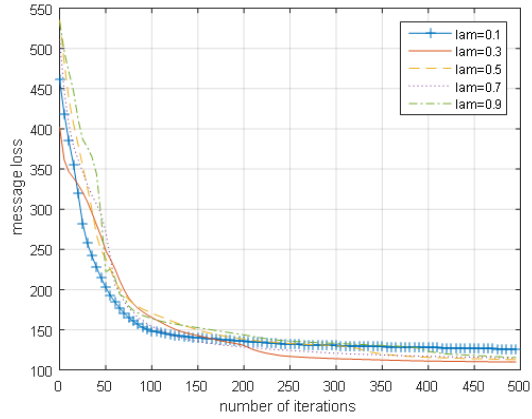

(b)

**FIGURE 9**. The message loss and the perceptual loss. (a) perceptual loss (b) message loss

We also present the results of completing the same image using different $\lambda$ values in Fig. 10. Although the gaps between the generated stego images are initially large, the completion stego images tend to become increasingly similar as the number of iterations increases.

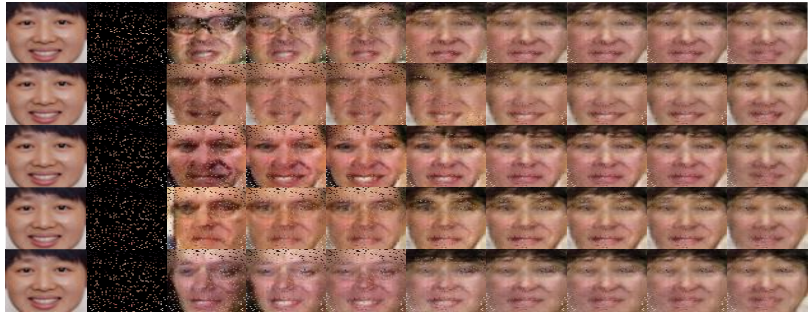

**FIGURE 10. Multiple stego images generated for a single image. In each row, column 1 shows ground-truth image, column 2 shows corrupted stego images with random missing regions comprising 90% of the image, and columns 3–10 show samples generated as the number of iterations increases.**

## C. QUANTITATIVE ANALYSIS

Due to the nonconvexity of the models in the training scheme, we cannot guarantee that the generator will converge to a model that can enable perfect recovery the secret message from the steganographic image . Fig. 11(a) and (b) shows the relationships between the message extraction accuracy rate and the number of iterations with LWF.

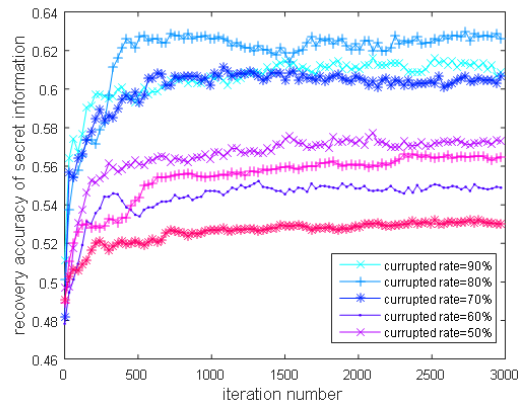

(a)

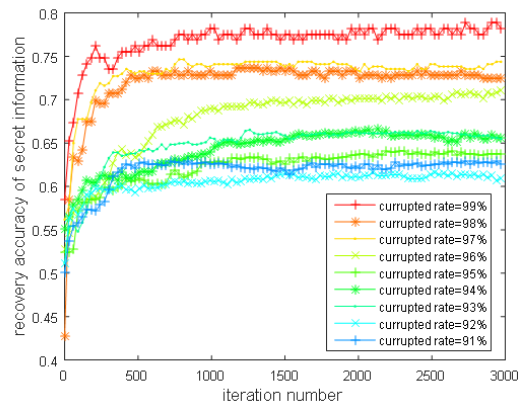

(b)

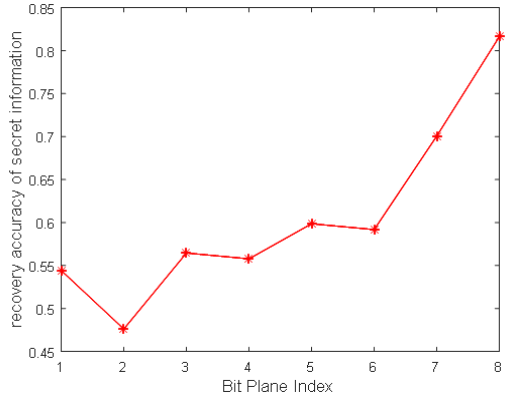

**FIGURE 11.** Message extraction accuracy rate for different corruption rate with BPI = 8  (a) 10%-90%, (b) 91%-99% (c) different BPI for   corruption rate 99% .

We choose to embed the message in the highest pixel of the uncorrupted image (bit plane index = 8), and then use the generator to sample stego images. The stego images are sampled every 30 iterations in a total of 3000 iterations. The recovery accuracy increases as the number of sampling iteration numbers increases. Due to the randomness of pixel generation and the fact that the message constraint cannot be completely satisfied. As can be seen from the Fig.11(a) and (b), as the amount of embedding increases, the accuracy of message extraction decreases. As shown in Fig. 11(c), we also performed message embedding and extraction average accuracy at different BPIs for 1,000 images that were not in the training set. All the stego images were sampled over 3,000 iterations, starting with corrupted images with missing 99% regions of the image. As expected, the accuracy of message extraction increased as the BPI value increased. The receiver was able to recover more than 70% at BPI = 7. Fig. 11(c) shows the relationship between the average error rate and the BPI. Compared with our previous work reported in [21], the message extraction stability is improved.

Currently, the steganography capacities of cover-synthesis methods are much lower than those of traditional embedding-based steganography methods. In our GSS method, message embedding is carried out on uncorrupted region of the image, and the message embedding is guaranteed by a message loss constraint, therefore the image has a high embedding capacity. We compare the steganography capacity of our method with other state-of-the-art cover selection and the cover-synthesis methods in Table 1, where the second column shows the absolute steganography capacity (steganography capacity per pixel), the third column gives the size of the stego image, and the last column is the relative steganography capacity (steganography capacity per pixel):

$$Relative\ capacity = Absolute\ capacity/Size\ of\ the\ image \qquad (28)$$

**TABLE 1. Steganography capacities of various methods.**

| Reference | Absolute Capacity (bytes/image) | Image Size | Relative Capacity (bytes/pixel) |
|---|---|---|---|
| [11] | 3.72 | ≥512×512 | 1.42e-5 |
| [33] | 1.125 | 512×512 | 4,29e-6 |
| [34] | 2.25 | 512×512 | 8.58e-6 |
| [10] | 64×64 | 800×800 | 6.40e-3 |

| [7] | 1535~4300 | 1024×1024 | 1.46e-3~4.10e-3 |
| [31] | ≥37.5 | 64×64 | 1.46e-3~4.10e-3 |
| **Ours** | **18.3-135.4** | **64×64** | **1.49e-3~1.10e-2** |

The relative capacity of our method is 1.49e-3~1.10e-2 bytes per pixel, where the lower and upper ends of this range correspond to image corrupted rates of 0.99 to 0.91, respectively, as shown in the last row of Table 1, In theory, the relative capacity of the proposed method can be higher, but under the message loss constraint in the message, as shown in Fig. 11, extraction accuracy will be very low, so the high embedding amount does not have practical value, the actual capacity should be calculated as:

$$Actual\ capacity = Relative\ capacity \times Extraction\ rate \quad (29)$$

This is mainly because the message constraint conditions are based on the pixel values themselves. Due to the randomness of pixel generation and the fact that the message constraint cannot be completely satisfied, the actual embedding capacity is reduced.

We also steganalyzed our GSS method using a blind steganalyzer in the spatial domain and an ensemble classifier. In this experiment, we used 686-dimensional SPAM features [25] and 5,404-dimensional SCRMQ1 features [26] with an ensemble classifier [27] implemented based on random forests. The decision threshold for each base learner was adjusted to minimize the total detection error under equal priors on the training set [27]:

$$P_E = \min \frac{1}{2}(P_{FA} + P_{MD}(P_{FA})), \quad (30)$$

where $P_{FA}$ and $P_{MD}$ are the probabilities of false alarms and missed detections, respectively. We adjusted the threshold to L/2 because in practice, $P_E$ is currently regarded as the standard for evaluating the accuracy of steganalyzers.

Unlike in traditional steganalyzers for the cover modification method, all 1,000 stego images and 1,000 normal images were generated over 1,000 iterations from corrupted images via the image inpainting process. The database was randomly divided into two halves: one for training and the other for testing. We averaged the performances over ten random splits. Fig. 12 is a plot showing the evolution of the testing error PE as a function of the payload from 0.01 bits per pixel (bpp) to 0.05 bpp when BPI = 7. For comparision, this plot also includes the corresponding results for HUGO [28] and HILL [29], which are considered to be advanced steganographic methods that minimize distortion by means of STC.

The results in Fig.12 show that steganography based on sampling can resist statistical steganographic analysis, mainly because the completed stego and normal images can be regarded as samples from the distribution $p_g$. There is no pairwise relationship between the features extracted from the normal cover image and the stego image. As shown in Fig. 12, our method performs competitively with HUGO [28] and HILL [29] for cases with a low embedding rate.

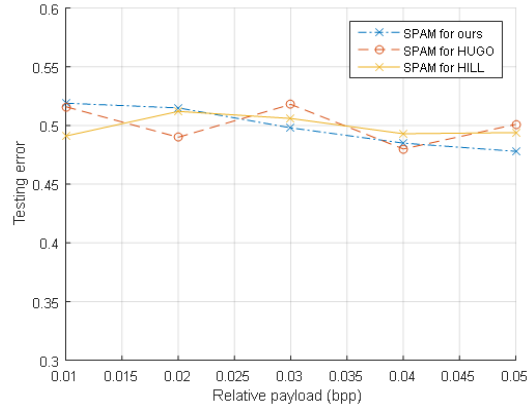

(a)

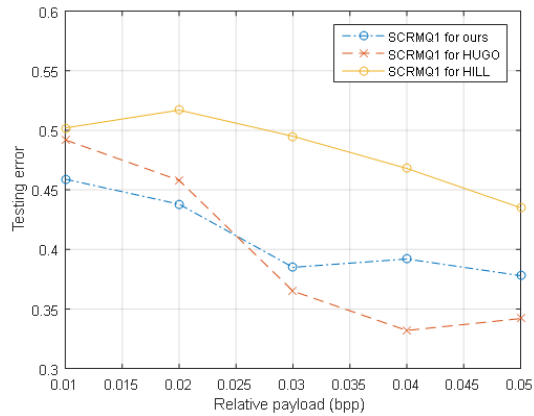

(b)

**FIGURE 12.** Steganalyzer error PE for an ensemble classifier using (a) SPAM features and (b) SCRMQ1 features for five different payloads with HUGO, HILL and our method.

## VI. CONCLUSIONS AND FUTURE WORK

In this paper, a generative steganography method is proposed in which the stego images are sampled by a well-trained generator, enabling them to resist statistical steganalysis. Inspired by the idea of the Cardan grille, we propose a practical method of generative steganography using image completion technology. The experimental results verify the promise of this simple method. It reduces the complexity of steganography design, allowing researchers in other fields to quickly build a steganographic system under this framework. However, the design of the generator, which is trained by an adversarial network, is still in its infancy. In this paper, we used a simple DCGAN for image synthesis. In future work, we will adopt a more powerful generator that will be able to automatically synthesizes more realistic images from either other images or text. The performance of the generator must be continually refined to ensure the security of generative steganography.